\documentclass[conference]{IEEEtran}
\IEEEoverridecommandlockouts
\usepackage{cite}
\usepackage{amsmath,amssymb,amsfonts}
\usepackage{algorithmic}
\usepackage{graphicx}
\usepackage{textcomp}
\usepackage{xcolor}
\usepackage{siunitx}
\usepackage{xcolor}

\def\BibTeX{{\rm B\kern-.05em{\sc i\kern-.025em b}\kern-.08em
		T\kern-.1667em\lower.7ex\hbox{E}\kern-.125emX}}
	\renewcommand{\baselinestretch}{0.95}
\begin{document}
	\onecolumn
	© 2022 IEEE.  Personal use of this material is permitted.  Permission from IEEE must be obtained for all other uses, in any current or future media, including reprinting/republishing this material for advertising or promotional purposes, creating new collective works, for resale or redistribution to servers or lists, or reuse of any copyrighted component of this work in other works.
	\newpage

	\twocolumn
	\title{	Impact of spatiotemporal heterogeneity in heat pump loads on generation and storage requirements}
	
	\author{\IEEEauthorblockN{Claire E. Halloran, Filiberto Fele, Malcolm D. McCulloch}
		\IEEEauthorblockA{\textit{Department of Engineering Science} \\
			\textit{University of Oxford}\\
			Oxford, United Kingdom \\
			\{claire.halloran, filiberto.fele, malcolm.mcculloch\}@eng.ox.ac.uk}
	}
	
	\maketitle
	\IEEEpubid{0000--0000/00\$00.00~\copyright~2022 IEEE}

	\begin{abstract}
	 This paper investigates how spatiotemporal heterogeneity in inflexible residential heat pump loads affects the need for storage and generation in the electricity system under business-as-usual and low-carbon emissions budgets. Homogeneous and heterogeneous heat pump loads are generated using population-weighted average and local temperature, respectively, assuming complete residential heat pump penetration. The results of a storage and generation optimal expansion model with network effects for spatiotemporally homogeneous and heterogeneous load profiles are compared. A case study is performed using a 3-bus network of London, Manchester, and Glasgow in Britain for load and weather data for representative weeks. Using heterogeneous heating demand data changes storage sizing: under a business-as-usual budget, 26\% more total storage is built on an energy and power basis, and this storage is distributed among all of the buses in the heterogeneous case. Under a low-carbon budget, total energy storage at all buses increases 2 times on an energy basis and 40\% on a power basis. The energy to power ratio of storage at each bus also increases when accounting for heterogeneity; this change suggests that storage will be needed to provide energy support in addition to power support for electric heating in high-renewable power systems. Accounting for heterogeneity also increases modeled systems costs, particularly capital costs, because of the need for higher generation capacity in the largest load center and coincidence of local peak demand at different buses. These results show the importance of accounting for heat pump load heterogeneity in power system planning.
	\end{abstract}
	

	\section{Introduction}
	This paper addresses the question: \textit{does weather-driven spatiotemporal heterogeneity in inflexible residential heat pump loads affect the need for storage and generation in the electricity system?} This question is answered with consideration of transmission constraints. The United Kingdom (UK) Climate Change Committee's Sixth Carbon Budget identifies residential building-scale low-carbon heat as a key area for greenhouse gas emissions reduction in buildings and foresees electrification of heating using heat pumps as the primary means for achieving this reduction \cite{CCC2020a}. Thus, heat pumps are poised to be a key decarbonization technology in the UK over the next 3 decades.
	
	One gap in the literature is the effect of electrified heating on the need for battery storage. A number of studies in the last decade focus on the system effects of widespread heat pump uptake and the role of heating flexibility in renewable integration \cite{Gaur2020,Guminski2019,Heinen2017}. However, these studies exclude non-heat forms of storage.
	
	Another gap in the literature is the effect of spatiotemporal heterogeneity in heating demand on electricity infrastructure requirements. Several studies have demonstrated that weather, housing stock, and other socio-economic factors will lead to spatial variation in electrified heating load \cite{Eggimann2019,White2021,Lombardi2022}. This weather-driven load heterogeneity may impact the power system infrastructure needed to integrate heat pumps; however, no study to date has explored this effect.
	
	This work addresses these gaps by assessing how spatiotemporal heterogeneity in heat pump load affects the need for electricity infrastructure, including battery storage sizing and siting. We utilize a storage and generation expansion problem formatted as a linear optimization. This model sizes and sites new generation and storage at multiple nodes considering transmission thermal limits. Based on data from heat pump trials, residential heating loads are calculated using outdoor air temperature: homogeneous loads are calculated using population-weighted average temperature for the grid region, and heterogeneous loads are calculated using local temperature. To test whether heterogeneous heating demand significantly impacts generation and storage planning, this model is applied to a case study of 3 urban areas in Britain. 
	
	\IEEEpubidadjcol
	\section{Methodology}

	\subsection{Heating demand model}
	To explore the effect of spatiotemporally heterogeneous heat pump demand on power system infrastructure requirements, we compare deterministic spatiotemporally heterogeneous heating demand, calculated based on local temperature, and deterministic spatiotemporally homogeneous heating demand, calculated based on population-weighted average national temperature. We use a model from Watson \textit{et al.} that generates heat pump electricity demand profiles as a function of the number of households using heat pumps, the share of air versus ground source heat pumps, and average daily outdoor air temperature \cite{Watson2021}. This model is chosen for its basis in heat pump trial data and nationally representative boiler gas demand data in Britain, the geography considered in the case studies. To establish the maximum possible impact of heat pump use on the power system, we assume that all households use heat pumps for space and water heating. Based on the air to ground source heat pump ratio observed in the trials, we assume that the heat pumps are 25\% ground source and 75\% air source \cite{Watson2021}.
	
	Homogeneous heating demand is calculated in the Watson \textit{et al.} model \cite{Watson2021} using a population-weighted mean daily temperature \cite{GMAO2008,Reis2017}. This national average daily temperature is used to calculate national heating demand $p_{heat}^{natl}$, which is then allocated to each area $i$ considered in the case studies based on their share of the national population $PS_i$, as shown in (\ref{eq:homogeneous heat demand}). Total power demand for each area is found by summing this value with $p_{base}$, deterministic non-heating power demand based on historical data, weighted by population share: 
	\begin{equation}\label{eq:homogeneous heat demand}
		p_{d,i}^{homo}(t)=PS_i \; p_{base}(t)+PS_i \; p_{heat}^{natl}(t)
	\end{equation}
	Heterogeneous heating demand for each area $p_{heat,i}^{local}$ is calculated from the Watson \textit{et al.} model \cite{Watson2021} with local mean daily temperature and the number of households in each area. The resulting total demand $p_{d,i}^{het}$ has a different load profile shape and magnitude in each location:
	\begin{equation}\label{eq:heterogeneous heat demand}
		p_{d,i}^{het}(t)=PS_i \; p_{base}(t)+p_{heat,i}^{local}(t)
	\end{equation}
	\subsection{Generation and storage expansion model}
	A generation and storage expansion optimization problem is used to meet total demand with the lowest possible operating and capital cost. The objective of this problem is minimizing system costs, that is, operating costs from all generator technologies $g$ at all buses $i$ for each time step $t$ considered, plus the capital costs of investments in new generation capacity and storage capacity.
	\begin{equation}\label{eq:objective}
	\begin{split}
		\min \sum_{g,i,t} c_{gen,g} \; p_{g,i}(t) \; \tau 
		+ \sum_{g,i}  c_{cap,g} \; C^{new}_{g,i}\\
		 + \sum_{i}  c_{SOC} \; SOC^{max}_i
		 + \sum_{i}  c_{cap,s} \; p^{max}_{s,i}
	\end{split}
	\end{equation}
	where $c_{gen,g}$ is the dispatch cost of each technology, $ c_{cap,g}$  is the new build capacity cost for each generation technology, and $c_{SOC}$ and $c_{cap,s}$ are the cost of energy and power storage capacity. The time duration between each time step $t$ is $\tau$. The decision variables are: $p_{g,i}(t)$, the power generated by each technology at each bus at each time, $C^{new}_{g,i}$, the new generation capacity build of each technology at each bus, $SOC_i^{max}$, the maximum state of charge in \si{\giga \watt \hour} (also known as energy capacity) for the storage unit at each bus, and $p^{max}_{s,i}$, the maximum storage power at each bus. The capital costs are divided quarterly over the expected lifetime of the assets since only 12 representative weeks are considered. 
	
	This objective function is subject to the constraints in \eqref{eq:bus-network power balance}-\eqref{eq:storage periodicity}. Unless otherwise noted, these equations apply for all buses $i$ in the set of buses $\mathcal{I}$, all times $t$ in the set of time steps in the time period considered $\mathcal{T}$, and for all generation technologies $g$ in the set of modeled generation technologies $\mathcal{G}$. Equation \eqref{eq:bus-network power balance} shows the bus-network power balance constraint. $L_{ij}$ is the bus-line matrix, which is equal to 1 if line $l$ originates at bus $i$, $-1$ if line $l$ terminates at bus $i$, and 0 otherwise. Positive flow $p_{line+,l}$ corresponds to flow in the same direction of the line, while negative flow $p_{line-,l}$ is in the opposite direction listed in the bus-line matrix.  $\mathcal{L}_i$ is the set of lines connected to bus $i$. The DC power flow approximation is \eqref{eq:DCPF}: $B_{ij}$ is the matrix of the reactance of the line between buses $i$ and $j$ and $\delta_i$ is the voltage angle at bus $i$.	Equation \eqref{eq:line limits} describes the thermal line limits, where $\mathcal{L}$ is the set of all lines in the network. Generation limits based on the availability of each generation technology are shown in \eqref{eq:generation limits} for the set of modeled generation technologies $\mathcal{G}$. The availability factor $A_{g,i}$ is a value between 0 and 1 that represents the location-specific availability of each generation technology. The availability factor for non-renewables generation technologies is always 1. The existing generation capacities are represented as $C^{old}_{g,i}$. Equation \eqref{eq:carbon budget} reflects the carbon budget limit on fossil fueled generator use, with $\epsilon_g$ as the carbon emissions intensity of generating a unit of energy from technology $g$, $\tau$ the length of time interval $t$, and $E_{lim}$  the emission limit set for time period $\mathcal{T}$. 
	\begin{equation}\label{eq:bus-network power balance}
		\begin{split}
			p_{net,i} (t)=
			\sum_{l} L_{il} (p_{line+,l} (t)-p_{line-,l} (t))
			\;\forall l\in \mathcal{L}_i
		\end{split}
	\end{equation}
	\begin{equation}\label{eq:DCPF}
		\begin{split}
			p_{line+,l} (t)-p_{line-,l} (t)=B_{ij} (\delta_i-\delta_j)\\
			\;\forall i,j \; \text{connected by line} \;l
		\end{split}
	\end{equation}
	\begin{equation}\label{eq:line limits}
		\begin{split}
			-p_l^{max}\leq p_{line+,l} (t)-p_{line-,l} (t)\leq p_l^{max}
			\; \forall l \in \mathcal{L}	
		\end{split}
	\end{equation}
	\begin{equation}\label{eq:generation limits}
		0\leq p_{g,i}(t)\leq A_{g,i}(t) (C^{old}_{g,i}+C^{new}_{g,i})
	\end{equation}
	\begin{equation}\label{eq:carbon budget}
		\sum_{g,i,t}  \tau \; \epsilon_g \; p_{g,i}(t) \leq  E_{lim} 
	\end{equation}
Equation \eqref{eq:bus-generator power balance} enforces the bus-generator power balance, where $p_{g2n,i}$ is the power from the generators injected into the network, $p_{g2d,i}$ is the power from the generators used to meet load, and $p_{g2s,i}$ is the power from the generators stored in local storage. Equation \eqref{eq:bus net power balance} gives the net power balance at each bus, where $\eta$ is the storage discharge efficiency, $p_{n2s,i}$ is the power from the network stored at bus $i$, and $p_{n2d,i}$ is the power from the network used to meet load at bus $i$. Demand-supply balance is guaranteed by \eqref{eq:demand-supply balance}, where $p_{d,i}$ is the demand at each bus $i$, $p_{g2d,i}$ is the power from generation at bus $i$ used to meet demand at the same bus, and $p_{s2d,i}$ is the power from storage used to meet demand at the same bus $i$. For homogeneous demand scenarios, the demand with homogeneous heating calculated in (\ref{eq:homogeneous heat demand}) is used for $p_{d,i}$ in (\ref{eq:demand-supply balance}); the demand calculated in (\ref{eq:heterogeneous heat demand}) is used in heterogeneous demand scenarios.
	\begin{equation}\label{eq:bus-generator power balance}
			\sum_{g} p_{g,i} (t) = p_{g2n,i} (t)+p_{g2d,i}(t)
			+p_{g2s,i} (t)
	\end{equation}
	\begin{equation}\label{eq:bus net power balance}
	\begin{split}
		p_{net,i} (t)=p_{g2n,i} (t)+\eta p_{s2n,i}(t)
		-p_{n2s,i} (t)-p_{n2d,i} (t)
	 \end{split}
 	\end{equation}
 	\begin{equation}\label{eq:demand-supply balance}
 		\begin{split}
 			p_{d,i} (t)=p_{g2d,i} (t)+\eta \; p_{s2d,i} (t)
 			+p_{n2d,i} (t)
 		\end{split}
 	\end{equation}
	Equations \eqref{eq:storage power charge} through \eqref{eq:storage ratio limits} constrain the storage units. Equations \eqref{eq:storage power charge} and \eqref{eq:storage power discharge} give the storage charging and discharging power limits, respectively. The state of charge of each storage unit is limited in \eqref{eq:storage SOC limits}), 	where $SOC_i$ is the state of charge of the storage unit at bus $i$. This is the state of charge from the previous time step plus the sum of power flows into and out of the storage unit at the previous time step, times the time step length $\tau$, as shown in \eqref{eq:storage integrator non-initial}. In the first time period considered, the state of charge is defined by the decision variable $SOC_{i,0}$ (see \eqref{eq:storage integrator initial}). Equation \eqref{eq:storage periodicity} enforces state of charge periodicity. Equation \eqref{eq:storage ratio limits} limits the ratio of storage energy capacity to power capacity to be between 1 and 4 hours, in line with typical values for grid-scale lithium-ion storage facilities.
	
	\begin{equation}\label{eq:storage power charge}
		0 \leq p_{g2s,i} (t)+p_{n2s,i} (t)\leq p_{s,i}^{max}
	\end{equation}
	\begin{equation}\label{eq:storage power discharge}
			0 \leq p_{s2n,i} (t)+p_{s2d,i} (t)\leq p_{s,i}^{max}
	\end{equation}
	\begin{equation}\label{eq:storage SOC limits}
				0\leq SOC_i (t) \leq SOC_i^{max}
	\end{equation}
	\begin{equation}\label{eq:storage integrator non-initial}
		\begin{split}
			SOC_i (t)=SOC_i (t-1)
			+p_{g2s,i} (t-1) \; \tau\\
			+p_{n2s,i} (t-1) \; \tau
			-p_{s2n,i} (t-1) \; \tau\\
			-p_{s2d,i} (t-1) \; \tau
			\;\forall \; t\in \mathcal{T}, t\neq0
		\end{split}
	\end{equation}
	\begin{equation}\label{eq:storage integrator initial}
		 SOC_i (t=0)=SOC_{i,0}
	\end{equation}
	\begin{equation}\label{eq:storage periodicity}
			SOC_i (t=final)=SOC_{i,0}
	\end{equation}
	\begin{equation}\label{eq:storage ratio limits}
		1 \leq SOC_i^{max}/p_{s,i}^{max} \leq 4
	\end{equation}
	
	\subsection{Case study: 3-bus Britain model}
 We consider a case study of 3-bus system in which each bus corresponds to an urban area in Britain: greater Glasgow, the Manchester built-up-area, and the London built-up-area. These urban areas were selected for their large populations as well as their large geographic distance from one another. 
	 
	Weather and electric demand data from representative weeks in 2019 are used. Non-heating electric demand data for each urban area was approximated by dividing the historical demand profile by the proportion of the population residing in each urban area \cite{NationalGridESO2020,UKOfficeforNationalStatistics2013,ScotlandsCensus2011}. This base demand was added to the heating demand profiles.
	
	Renewable potential profiles for different locations were generated based on historical weather data using the renewables.ninja tool \cite{Pfenninger2016}. For offshore wind potentials, historical weather data was taken from the location of the nearest offshore wind farm built by the end of 2018 (London Array for London, Walney for Manchester, and Beatrice for Glasgow). Because most load is often within urban areas and most generation is outside of urban areas, the renewable and conventional generation capacity of the entire NUTS 1 region\footnote{Nomenclature of Territorial Units for Statistics. NUTS 1 defines regional level subdivision of the UK territory.} at the end of 2018 was allocated to the corresponding urban area \cite{BEIS2020,DUKES2020}. All conventional generation capacity is modeled as combined cycle gas turbines (CCGT).	The London region has relatively little generation capacity, so half the capacity of South East England is assumed to be used to meet the electricity demand of the London built-up area. 
	
	Nuclear generation and interconnector imports are fixed at their historical levels for 2019 \cite{Exelon2019}. Grid-wide nuclear dispatch is allocated to each bus in proportion to the nuclear capacity in each region and priced at its marginal operation price of \$29,000 per GWh \cite{LazardEnergy2019}. Interconnector imports are allocated to the nearest bus considered in the case study and priced at \$11,000 per GWh, based on typical 2019 auction prices \cite{JAO2019}. Nuclear and interconnector capacity expansion are not considered in this case study.
	
	Representative weeks were selected for each month to capture low-wind, low-temperature extremes. Heinen \textit{et al.} identified these weather conditions as critical situations in systems with electrified heating and large wind generation in the context of Ireland \cite{Heinen2017}. Thus, for each month, minimum net demand hour is identified. Net demand is defined total wind generation based on 2018 wind capacity minus total heterogeneous demand. The sample week was the 4 days prior to the day on which this minimum net demand hour occurred and the following 3 days. This process results in a set of 12 sample weeks of weather and demand conditions that represent system extremes throughout the year. Capturing these extremes is crucial for ensuring that adequate capacity is built.

	Levelized cost of energy values for 2019 were used for the dispatch prices of all generation technologies, and dispatch from storage was assumed to have no cost \cite{LazardEnergy2019}. Capital costs for generation technologies were determined using quarterly nameplate capacity costs for new generation built in 2019 \cite{LazardEnergy2019}. For storage, 2019 nameplate energy capacity cost for lithium-ion batteries was used \cite{LazardStorage2019}. To minimize unused storage power capacity, an arbitrarily small price of \$10 per GW was assigned. A storage efficiency of 90\% is assumed.

	Transmission lines were assumed to connect all cities to one another. Line reactance values were calculated using the distance between cities and a per length reactance value of 0.019 \% p.u. per \si{km}, which is in line with the reactance values for high-voltage transmission lines in the Great Britain system \cite{ETYS2020}. Transmission capacity between each bus was assumed to be \SI{5}{\giga \watt}.
	
	This case study was performed with two different carbon budgets based on the UK Climate Change Committee's (CCC) Sixth Carbon Budget for electricity generation \cite{CCC2020}: a 2019 budget reflecting business-as-usual emissions and a 2035 budget reflecting aggressive decarbonization. Because the urban areas included represent about one-fifth of the total UK population and the high-demand winter period is considered, 10\% of the CCC carbon budget is used for this case study. A carbon intensity of \SI{365}{\gram} CO$_2$e per kWh \cite{POST2011} is assumed for CCGT generation. All other generation is considered carbon-free.
	
	\section{Results and discussion}

	\subsection{Homogeneous and heterogeneous heat demand}
	In the case study considered, using a single grid-wide temperature to generate homogeneous demand profiles did not fully capture the high heat demand periods at each bus that are present in the heterogeneous demand profiles. Regardless of the heating profile used, this model suggests that switching to heat pumps for domestic space and water heating will drastically increase electric demand in urban areas. However, using a local temperature-based model leads to much higher peak and average demand: for instance, considering homogeneous heating demand about doubles the historical average demand in London to \SI{11}{\giga \watt} and heterogeneous demand further increases average demand to \SI{13}{\giga \watt}. Peak demand for the London urban area also increases from \SI{22}{\giga \watt} in the homogeneous case to \SI{30}{\giga \watt} in the heterogeneous case. The homogeneous demand profile based on national population-weighted average temperature data, including warmer coastal areas in the south, lacks the peaks and valleys in heating demand that are clear in the heterogeneous profiles of these inland cities. For the winter peak demand period considered, the differences in demand can be relatively large, and capturing these extreme grid conditions is crucial for power system planning. 
	
	\subsection{3-bus Britain model}

\begin{figure}
	\begin{centering}
	\includegraphics[width=0.95\columnwidth]{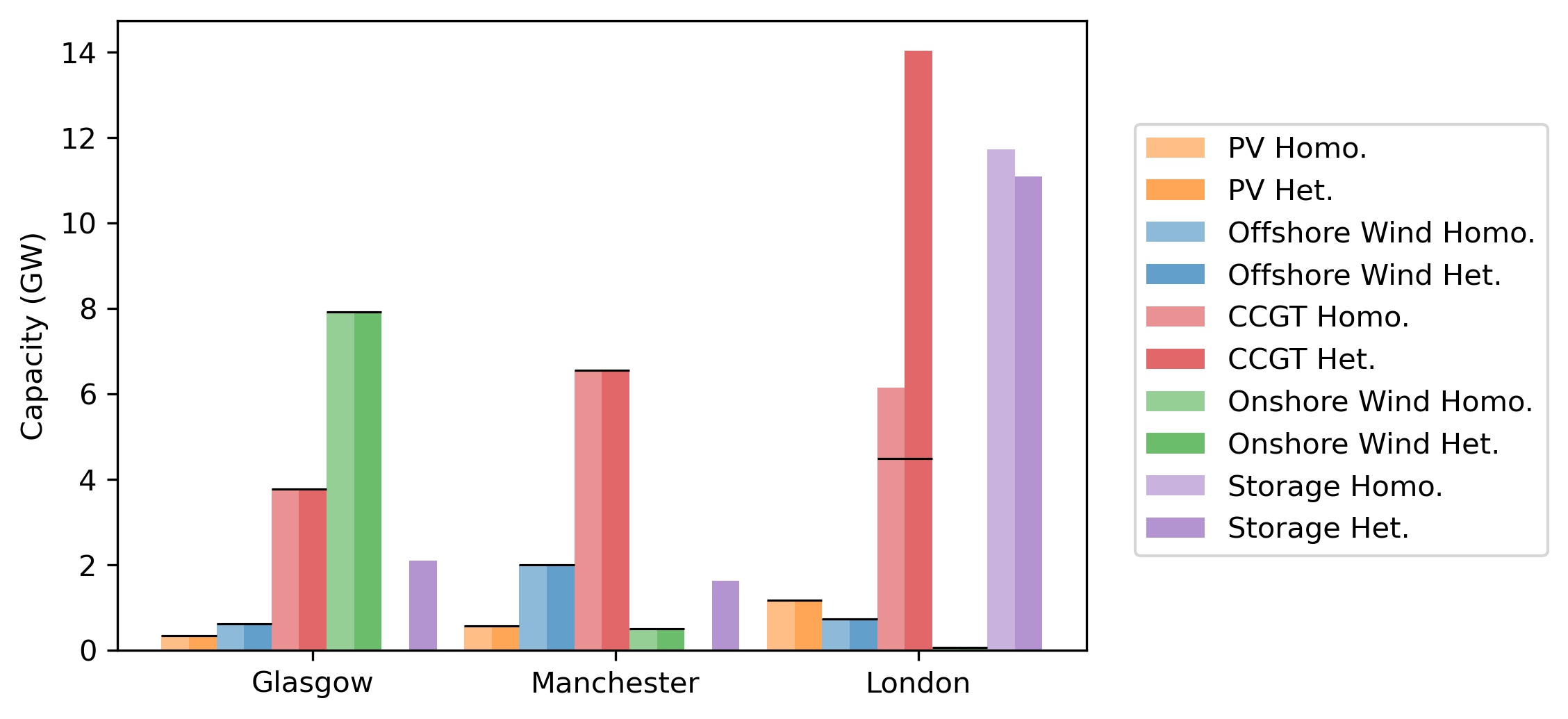}
	\end{centering}
	\caption{Capacity expansion for homogeneous and heterogeneous heating demand under 2019 carbon budget, with existing generation capacity in 2018 indicated by the black line.}
	\label{fig:CapacityExpansion2019}
\end{figure}

	\subsubsection{2019 Carbon Budget}
	Figure \ref{fig:CapacityExpansion2019} shows the added capacity of different technologies at each bus to meet electric  heating demand for a 2019 carbon budget. 	Storage capacity is added at all buses to meet heterogeneous heating demand, but the storage capacity at the London bus is slightly decreased for heterogeneous demand.  Only the London bus adds any generation capacity, all of which is CCGT, to meet both homogeneous and heterogeneous demand. Because transmission constraints limit power imports from other buses, the London bus requires additional generation to meet its large peak demand. Added CCGT capacity increases dramatically from \SI{1.7}{\giga \watt} in the homogeneous case to \SI{9.6}{\giga \watt} in the heterogeneous case. This fivefold increase is much larger than the 36\% increase in peak demand between the two  cases because in the heterogeneous case high demand in London coincides with high demand in Manchester on the maximum CCGT use day. This coincidence necessitates more local CCGT capacity in London to meet demand since imports from other buses cannot be relied upon. 
	
	\begin{figure}
		\begin{center}
		\includegraphics[width=0.85\columnwidth]{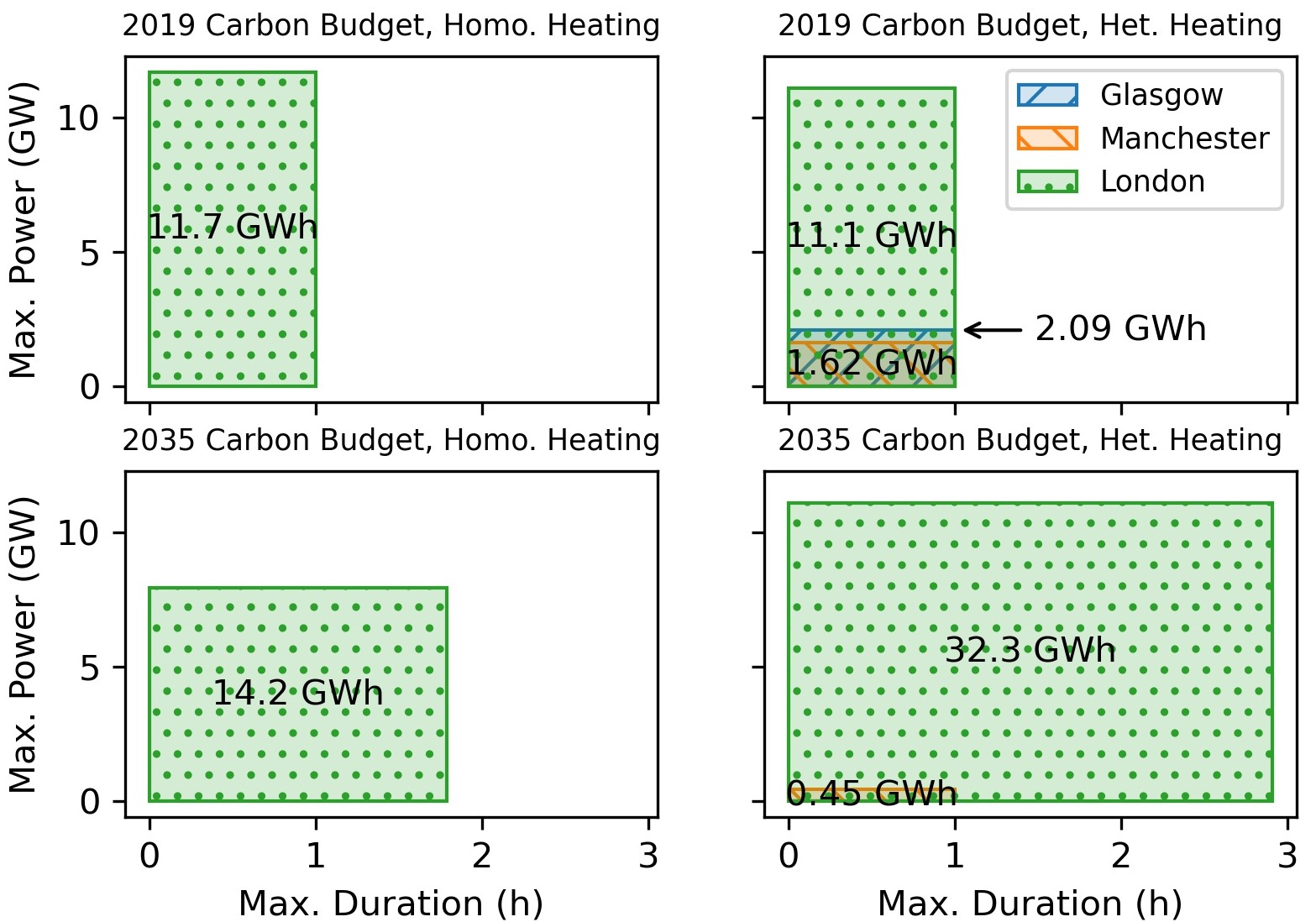}
		\end{center}

		\caption{Storage size at each bus, indicated by the rectangle color. The area of each rectangle is the energy capacity.}
		\label{fig:StorageSize}
	\end{figure}
	
	More detail about storage size and distribution is shown in Figure \ref{fig:StorageSize}. Storage is only deployed at the London bus in the homogeneous demand case, but is distributed among all buses in the heterogeneous demand case. The total storage deployed also slightly increases from  \SI{11.7}{\giga \watt \hour} to \SI{14.8}{\giga \watt \hour}. Storage power and energy capacity at the London bus slightly decreases from \SI{11.7}{\giga \watt} in the homogeneous case to \SI{11.1}{\giga \watt} in the heterogeneous case because the CCGT capacity added at the bus can partially substitute for storage.
		
\begin{figure}
	\begin{center}
	\includegraphics[width=0.7\columnwidth]{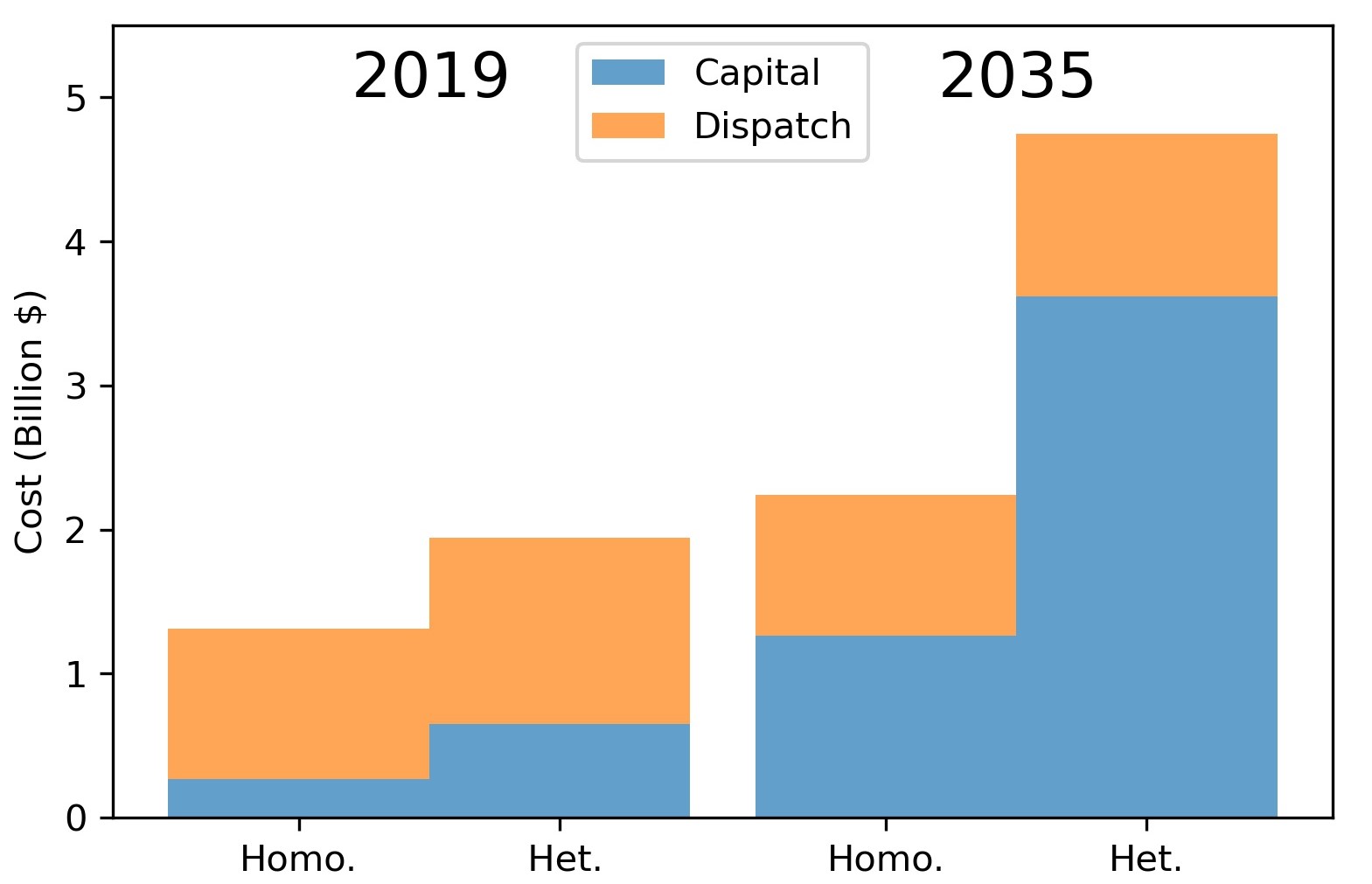}
	\end{center}

	\caption{System capital and dispatch costs under each scenario.}
	\label{fig:SystemCosts}
\end{figure}

	The impact of considering heating demand heterogeneity on estimated systems costs can be observed in Figure \ref{fig:SystemCosts}, which shows the quarterly capital and dispatch costs. Heterogeneous heating load results in a 2.4 times higher capital cost than homogeneous demand. This difference is primarily driven by the higher CCGT capacity deployed at the London bus and increase in total storage capacity. A much smaller difference is observed in the dispatch cost for the two load scenarios: meeting heterogeneous load is 14\% more expensive than meeting homogeneous heating demand, mostly due to the additional energy needed to supply higher average demand.

\begin{figure}
	\begin{centering}
	\includegraphics[width=0.95\columnwidth]{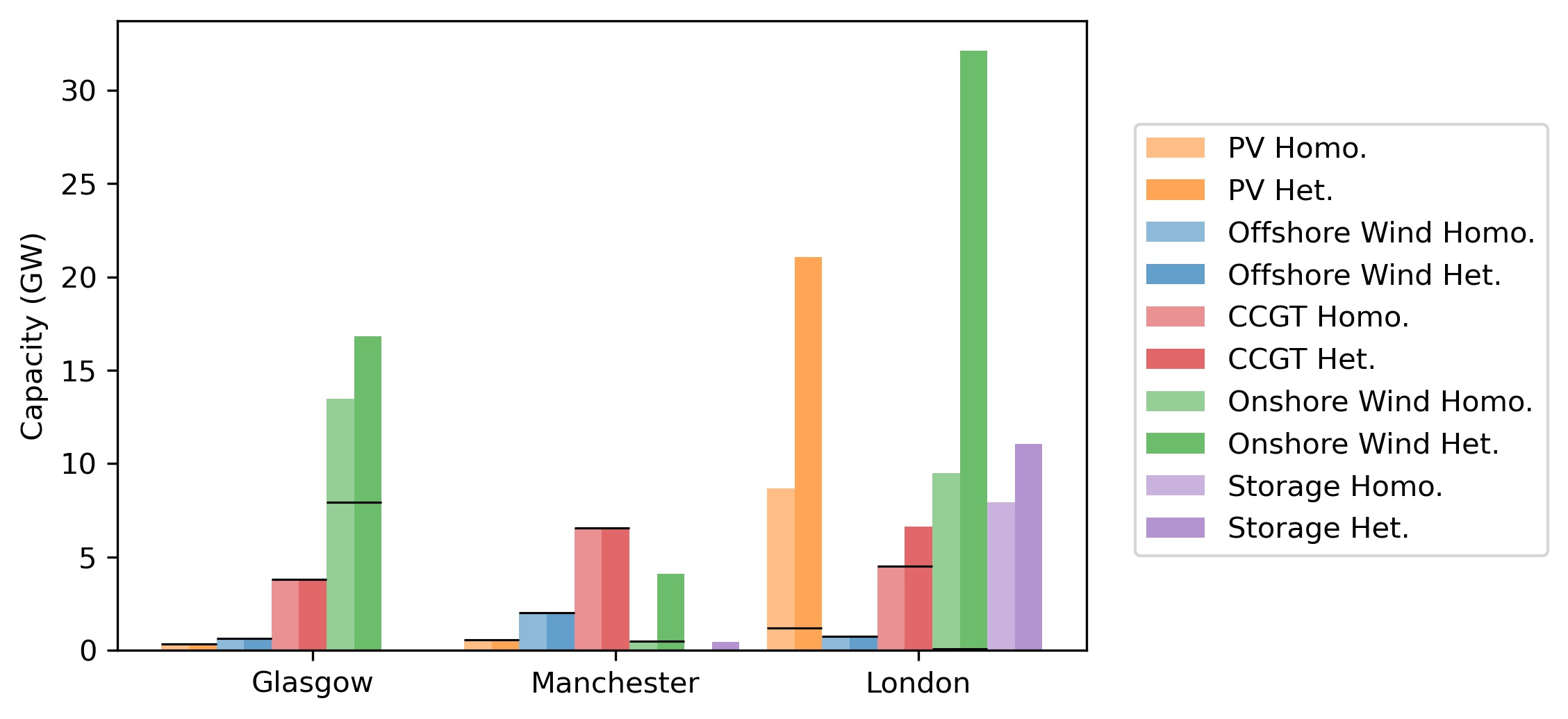}
	\end{centering}
	\caption{Capacity expansion for homogeneous and heterogeneous heating demand under 2035 carbon budget, with existing generation capacity in 2018 indicated by the black line.}
	\label{fig:CapacityExpansion2035}
\end{figure}

	\subsubsection{2035 Carbon Budget}
	Fig. \ref{fig:CapacityExpansion2035} shows capacity expansion at each bus considering homogeneous and heterogeneous heating demand under the 2035 carbon budget. In the heterogeneous case, additional onshore wind is added at all buses to compensate for the limit on CCGT generation and storage is added at the London and Manchester buses. 
	
	As in the 2019 carbon budget, higher peak load under the heterogeneous scenario in combination with transmission constraints drives drastic generation expansion in London. Due to the strict carbon budget in 2035, the majority of this capacity is onshore wind and photovoltaic (PV), though some additional CCGT is built. Onshore wind capacity in London over triples from \SI{9}{\giga \watt} in the homogeneous case to \SI{32}{\giga \watt} in the heterogeneous case. PV capacity increases by 2.7 times from \SI{7.5}{\giga \watt} in the homogeneous case to \SI{19.9}{\giga \watt} in the heterogeneous case.
	 	
	Figure \ref{fig:StorageSize} also shows storage sizing and distribution added under the 2035 carbon budget. Compared to the 2019 carbon budget scenario, more total storage energy capacity but less total storage power capacity is deployed under the 2035 carbon budget. This increased maximum duration suggests that storage is used to provide energy for longer periods of time than under the 2019 budget, probably due to partial substitution for CCGT flexible generation. Large differences between storage expansion in the homogeneous and heterogeneous case are notable. Storage energy capacity increases by 2.3 times in the heterogeneous heating scenario compared to the homogeneous case. The bulk of this energy capacity is added at the London bus to increase utilization of power from the larger onshore wind and PV capacity in the heterogeneous case. These results suggest failing to account for spatiotemporal heat pump load heterogeneity may lead to substantial underestimation of flexibility requirements for integrating heat pumps into a renewable-dominated electric system.
	
	Figure \ref{fig:SystemCosts} shows similar cost differences between homogeneous and heterogeneous heating demand for the 2019 and 2035 carbon budget. Quarterly capital costs are 2.9 times higher to meet heterogeneous heating load compared to homogeneous heating load. This difference is primarily driven by the large increase in  onshore wind capacity in London. Dispatch costs are 15\% higher for the heterogeneous case; this difference reflects the cost of meeting greater total energy demand. Quarterly capital costs for the same heating scenario are 4.6 and 5.6 times higher for the 2035 carbon budget than the 2019 carbon budget (for the homogeneous and heterogeneous scenario respectively). This cost increase reflects the cost of building oversized intermittent renewable generation to decrease the use of CCGT generation during low-wind periods. 

	\section{Conclusion}
	This paper explored whether spatiotemporal heterogeneity in residential electricity demand under 100\% heat pump penetration affects power system generation and storage requirements. In the case study considered, using local temperature-based, heterogeneous heating demand leads to 18\% higher average demand and 36\% higher peak demand in London, the largest load center.  Because of transmission constraints, the PV and onshore wind capacity nearly triples in London to meet higher peak demand in the heterogeneous case under a strict carbon budget, compared to the homogeneous case. Coincident peaks in heterogeneous demand at London and Manchester also contribute to the need for higher generation capacity. Storage energy capacity also doubles from \SI{14}{\giga \watt \hour} to \SI{33}{\giga \watt \hour} under a strict carbon budget, and storage is sited at multiple locations in the heterogeneous case for both carbon budgets instead of only in London. These generation and storage increases lead to over double capital costs under a business-as-usual carbon budget and nearly triple capital costs under a strict carbon budget. These results suggest that failing to account for weather-driven spatiotemporal heat pump load heterogeneity in power system planning studies will lead to underestimation of storage and generation requirements and of system costs for heat pump integration. The substantial differences in this illustrative case study indicate that heterogeneity will influence power system planning in larger networks in future work. Future studies may also consider how weather-driven spatiotemporal heterogeneity in heating demand influences flexible heat pump operation. Another area for further investigation is incorporating uncertainty in weather conditions by formatting this planning problem as a two-stage optimization with multiple possible operation scenarios.
	
	\bibliographystyle{plain}
	\bibliography{sources}

\end{document}